\documentclass[a4paper,10pt]{article}
\usepackage{amsmath}
\usepackage{comment} 
\usepackage[pdftex]{graphicx}
\usepackage{cite}
\usepackage{authblk}

\usepackage{longtable}
\usepackage{multirow}

\usepackage{amsfonts}		
\usepackage{array}
\usepackage{graphicx}
\usepackage{caption}
\usepackage{subcaption}
\usepackage[utf8]{inputenc}
\usepackage{float}
\title{Neutrino masses, mixing, and leptogenesis in an S3 model}
\author{Arturo \'Alvarez Cruz and Myriam Mondrag\'on\\
\small Instituto de F\'isica, Universidad Nacional Aut\'onoma de M\'exico\\
Apdo. Postal 20-364, M\'exico 01000 CD MX, M\'exico}
\begin{document}
\maketitle
\begin{abstract}

  In this work we use previous results on the masses and mixing of
  neutrinos of an S3 model with three right-handed Majorana neutrinos
  and three Higgs doublets, to reduce one parameter in the case when
  two of the right-handed neutrinos are mass degenerate. We derive a new
  parameterization for the $V_{PMNS}$ mixing matrix, with a new set of
  parameters, in the more general case where the right-handed
  neutrino masses are different.  With these results, we calculate
  leptogenesis and the associated baryogenesis in the model in the two
  different scenarios.  We show that it is possible to have enough
  leptogenesis to explain the baryonic asymmetry with right-handed
  neutrino masses above $10^6$ GeV.

\end{abstract}

\section{Introduction}
The Standard Model (SM) is extremely successful, nevertheless the
discovery of neutrino masses and mixing in neutrino oscillation
experiments in 1998 \cite{Fukuda:1998mi}, presented evidence that is
necessary to go beyond it. Even before this discovery, the amount of
free parameters and the hierarchy problem, among others, have prompted
attempts to find a more fundamental theory, of which the SM is the
low-energy limit \cite{Ehrenfeld:2008hc,Georgi:1974sy,Georgi:1975qb}.
Some of the goals of these new models are to understand the large
differences in the Yukawa couplings of the different fermions, the
hierarchy between the fundamental particles, and the amount of CP
violation and the structure of the CKM matrix \cite{Masiero:2005ua}.
A popular way to approach these problems is to build models with
Non-Abelian flavor symmetries, often supplemented with extra Higgs
doublets. Common symmetries in flavor theories are, among many others,
A4, Q6 or S3
\cite{Felipe:2013ie,Felipe:2013vwa,Gomez-Izquierdo:2013uaa,Ishimori:2010au,Kubo:2003iw,Mondragon:2007jx}.
The reason is that these models achieve in a natural way the Nearest
Neighbour Interaction textures in the fermion mass matrices
\cite{Harayama:1996am, Harayama:1996jr}.  The S3 extension of the SM
with three Higgs doublets (S3-3H)
\cite{Kubo:2003iw,Mondragon:2007jx,Felix:2006pn} is a model in which a
symmetry on the permutation of three objects is imposed, which in
additon to the SM particles has another two Higgs doublets, as well as
three right-handed Majorana neutrinos, which are related to the left
ones through the seesaw mechanism (type I).

There has been a lot of work done on various S3 models (see for instance
\cite{Kubo:2004ps,Mondragon:2007nk,Mondragon:2007af,Lavoura:2005kx,Cogollo:2016dsd,Chen:2007zj,Dev:2011qy,Grimus:2005mu}),
some of this work reproduces the CKM and PMNS matrices in agreement
with the current experimental data
\cite{Canales:2012ix,Beltran:2009zz,Canales:2013cga,Mondragon:2007jx,Mondragon:2008gm,Kubo:2003iw},
and there have been also studies of leptogenesis in a soft breaking S3
model \cite{Araki:2005ec}.  Nevertheless, most of this work has been
done in the case where two right-handed neutrinos are degenerate.  In
this way, it is an interesting question to extend the model and see
the possible new results with a generalization, taking into account
both degenerate and non-degenerate right-handed neutrino masses.
Following the idea of previous work \cite{Canales:2012dr}, we extend the analysis on the generalization of the S3-3H model.\\

Another question that the  SM  fails to explain is the observed baryon asymmetry.
It is well know that there are more baryons than antibaryons in the Universe. Nucleosynthesis is a solid and  consistent 
model of the creation of the nuclei in the early Universe, which
predicts a baryonic density of, 
\begin{equation}
\eta=\frac{\eta_b-\eta_{\overline{b}}}{\eta_\gamma}=\eta=(2.6-6.2)\times10^{-10}.
\end{equation}
Measurements of the Cosmic Background Radiation  \cite{Steigman:1976ev,Trodden:2004st,Bennett:2003ca} show a density of   
\begin{equation}
\eta=(6.1\pm0.3)\times10^{-10},
\end{equation}
in full agreement with the baryon density of the Nucleosynthesis\cite{Lyth:1186232,Trodden:2004st}.\\
The idea to explain the baryon asymmetry through a dynamically process
was proposed by Sakharov in 1967 \cite{Sakharov:1967dj}.   The present 
cosmological observations favour the idea  that the matter-antimatter asymmetry
of the Universe may  be explained in terms of a dynamical generation
mechanism, called baryogenesis.  Also, it has been realized that a
successful model of baryogenesis cannot occur within the Standard
Model (SM).\\

Leptogenesis is a mechanism which generates baryon asymmetry by
creating a leptonic asymmetry through B + L violating electroweak
sphaleron transitions \cite{Kuzmin:1985mm}.
\\
Several things are needed for the occurrence of leptogenesis:
\begin{itemize}
  \item Heavy right handed neutrinos.
  \item Majorana type neutrinos.
  \item Decay of the right handed neutrinos to the left ones.
\end{itemize}
According to the original proposal of Fukugita and Yanagida
\cite{Fukugita:1986hr}, this mechanism also satisfies all the
Sakharov’s conditions \cite{Sakharov:1967dj} in order to produce a net
baryon asymmetry (for reviews see for instance \cite{Petcov:2016ovu,Molinaro:2015dta,Davidson:2008bu}).

In this paper we explore the possibility of leptogenesis in the S3-3H
model, with degenerate and non-degenerate right-handed neutrino
masses, and calculate the associated baryogenesis.  We first study the
case where two of the right-handed neutrino masses are degenerate, and
then the more general case where all the right-handed neutrino masses
are different.  We scan the parameter space to find the leptogenesis
and associated baryogenesis dependence on the free parameters of the
model.  We find that there is a region of parameter space where enough
baryogenesis is produced through leptogenesis to explain the
baryon asymmetry of the Universe.\\
The outline of the paper is organized as follows: In section 2, the S3
model is introduced as well as some of its most important results. In
section 3 it is shown how to produce leptogenesis in the S3-3H model, and
the resultant baryogenesis is also computed. At the end, in section 4,
we conclude summarizing our main results.

\section{S3-3H model}

In the Standard Model analogous fermions in different generations have
identical couplings to all gauge bosons of the strong, weak, and
electromagnetic interactions \cite{Mondragon:2007nk}. 
The group S3 consists of the six possible
permutations of three objects $(f_1,f_2,f_3)$, and is the smallest
discrete non-abelian group. It has one 2-dimension irreducible
representation (irrep) and two of 1-dimension
\begin{equation}
F_s=(f_1,f_2,f_3) ,\quad  F_{d1}= - f_1 - f_2  + 2f_3 ,
\end{equation}
\begin{equation}
 f_{d2}= f_2 - f_3 .
\end{equation}
We can associate the particles in the model to doublets or to singlets with the following rules.
The direct product of two doublets $p_D^T = (p_{D1} , p_{D2} )$ and $q_D^T = (q_{D1} , q_{D2} )$ may be decomposed
into the direct sum of two singlets $r_s$ and $r_{s'}$ , and one doublet $r_D^T$ where
\begin{equation}
r_s=p_{D1}q_{D1}+p_{D2}q_{D2}\quad r_{s'}=p_{D1}q_{D2}-p_{D2}q_{D1}
\end{equation}
\begin{equation}
r_D^T=(r_{D1},r_{D2})=(p_{D1}q_{D2}+p_{D2}q_{D1},p_{D1}q_{D1}+p_{D2}q_{D2}).
\end{equation}
Since the Standard Model has only one Higgs $SU(2)_L$ doublet, which can only be an $S_3$
singlet, it gives mass to the particles in the $S_3$ singlet representation.
To give mass to the rest of the particles we extend the Higgs sector
of the theory, by adding two more Higgs doublets. 
The quark and Higgs fields are
\begin{equation}
Q^T=(u_L,d_L),u_r,d_r,
\end{equation}
\begin{equation}
L^t=(\nu_L,e_L),e_R,\nu_R \quad \text{and}\quad H.
\end{equation}
 All of the fields have three species, and we assume that each one
forms a reducible representation $1S \oplus 2$. The first two
generations will be assigned to the doublet S3 irrep, and the third
generation to the singlet.  This applies to quarks, leptons, Higgs
fields, 
and right-handed neutrinos. The doublets carry capital indices $I$ and $J$, which
run from 1 to 2, and the singlets are denoted by $Q_3 , u_{3R} , d_{3R} , L_{3} , e_{3R} , \nu_{3R}$ and $H_S$.
 The subscript 3 denotes the singlet representation and not the third generation. 
The most
general renormalizable Yukawa interactions of this model are given by \cite{Kubo:2003iw}
\begin{equation}L_Y=L_{Y_D}+L_{Y_U}+L_{Y_E}+L_{Y_{\nu}}\end{equation}\\[10pt]
where
\begin{equation} \label{eq:3}
 \begin{split}
 L_{Y_{D}}= & -Y^{d}_l\overline {Q_I}H_S d_{IR} - Y^{d}_3\overline{Q_3}H_S d_{3r}\\
 & -Y^{d}_2[\overline {Q_I}\kappa_{IJ} H_l d_{JR} - \overline{Q_I}\eta _{IJ} H_2 d_{JR}]\\
 & -Y^d_4\overline{Q_3}H _Id _{IR}   -   Y^d_5\overline{Q_I}H _ID _{3R}  + h.c.  
 \end{split}
\end{equation}
\\
\begin{equation}
 \begin{split}
L_{Y_{U}}=&-Y^{u}_1\overline{Q_I}(i\sigma _2 H^{\ast} _Su _{IR})   -   Y^{u}_3\overline{Q_3}(i\sigma _2 H^{\ast} _Su _{3R})\\
&-Y^{u}_2[\overline{Q_I}\kappa_{IJ}(i\sigma _2 H^{\ast} _1u _{JR})  -  \overline{Q_I}\eta_{IJ}(i\sigma _2 H^{\ast} _2u _{JR})]\\
&-Y^u_4\overline{Q_3}(i\sigma _2 H^\ast _Iu _{IR})   -   Y^u_5\overline{Q_I}(i\sigma _2 H^\ast _Iu _3R)  + h.c.
 \end{split}
\end{equation}
\\
\begin{equation}
 \begin{split}
L_{Y_{U}}=&-Y^{u}_1\overline{Q_I}(i\sigma _2 H^{\ast} _Su _{IR})   -   Y^{u}_3\overline{Q_3}(i\sigma _2 H^{\ast} _Su _{3R})\\
&-Y^{u}_2[\overline{Q_I}\kappa_{IJ}(i\sigma _2 H^{\ast} _1u _{JR})  -  \overline{Q_I}\eta_{IJ}(i\sigma _2 H^{\ast} _2u _{JR})]\\
&-Y^u_4\overline{Q_3}(i\sigma _2 H^\ast _Iu _{IR})   -   Y^u_5\overline{Q_I}(i\sigma _2 H^\ast _Iu _3R)  + h.c.
 \end{split}
\end{equation}
\\
\begin{equation}
 \begin{split}
L_{Y_{E}}=&-Y^{e}_1\overline{L_I}H _Se _{IR})   -   Y^{e}_3\overline{L_3}H _Se _{3R})\\
&-Y^{e}_2[\overline{L_I}\kappa_{IJ}H_1e _{JR}  -  \overline{L_I}\eta_{IJ}H _2e _{JR})]\\
&-Y^e_4\overline{L_3}H _Ie _{IR}   -   Y^e_5\overline{L_I}H _ID _{3R}  + h.c.
 \end{split}
\end{equation}
\\
\begin{equation} \label{eq:4}
 \begin{split}
 L_{Y_{\nu}}=&-Y^{{\nu}}_1\overline{L_I}(i\sigma _2 H^{\ast} _S{\nu} _{IR})   -   Y^{{\nu}}_3\overline{L_3}(i\sigma _2 H^{\ast} _S{\nu} _{3R})\\
&-Y^{{\nu}}_2[\overline{L_I}\kappa_{IJ}(i\sigma _2 H^{\ast} _1{\nu} _{JR})  -  \overline{L_I}\eta_{IJ}(i\sigma _2 H^{\ast} _2{\nu} _{JR})]\\
&-Y^{\nu}_4\overline{L_3}(i\sigma _2 H^\ast _I{\nu} _{IR})   -   Y^{\nu}_5\overline{L_I}(i\sigma _2 H^\ast _I{\nu} _3R)  + h.c.,
 \end{split}
\end{equation}
with
\begin{equation} \kappa=\begin{pmatrix}
 0 & 1\\
 1 & 0\\
\end{pmatrix}
 \eta= \begin{pmatrix}
 1 & 0 \\ 
 0 & -1 \\
\end{pmatrix}.
\end{equation}
Furthermore, we add to the Lagrangian the Majorana mass terms for the right-handed neutrinos
\begin{equation} \label{eq:5}
L_M = -M_1\nu ^T_{1R}C\nu_{1R} -M_2\nu ^T_{2R}C\nu_{2R} -M_3\nu ^T_{3R}C\nu_{3R}. 
\end{equation}
Due to the presence of three Higgs fields, the Higgs potential $V_H (H_S , H_D )$ is more 
complicated than that of the Standard Model \cite{Pakvasa:1977in}. In addition to the $S3$
symmetry, under certain conditions  the Higgs potential
exhibits a permutational symmetry
$Z2 : H1 \leftrightarrow H2$, which is not a subgroup of the flavor
group S3 \cite{Beltran:2009zz,Canales:2013cga}. The model has as well  an Abelian discrete symmetry
 that we will use as selection rules for the Yukawa couplings in the
 leptonic sector. In this paper, we will assume
 that the vacuum respects the accidental $Z2$ symmetry of
the Higgs potential and that
\begin{equation}
<H_1>=<H_2>.
\end{equation}
With these assumptions, the Yukawa interactions, eqs. (\ref{eq:3})-(\ref{eq:4}) yield mass matrices, for
all fermions in the theory, of the general form
\begin{equation}
M=
\begin{pmatrix}
\mu_1+\mu_2 &\mu_2&\mu_5\\
\mu_2&\mu_1-\mu_2&\mu_5\\
\mu_4&\mu_4&\mu_3\\
\end{pmatrix}.
\end{equation}
The Majorana mass for the left handed neutrinos $\nu_L$ is generated by the see-saw mechanism.
The corresponding mass matrix is given by
\begin{equation}
M_\nu=M_{\nu D} \widetilde{M}^{-1}(M_{\nu D})^T,
\end{equation}
where  $\widetilde{M} = \text{diag}~(M_1 , M_1 , M_3 )$.
In principle, all entries in the mass matrices can be complex since there is no restriction coming
from the S3 flavor symmetry.
The mass matrices are diagonalized by bi-unitary transformations as
\begin{equation}
U^\dagger_{d(u,e)L}M_{d(u,e)}U_{d(u,e)R}=\text{diag}(m_{d(u,e)},m_{s(u,e)},m_{b(u,e)})
\end{equation}
\begin{equation}
U_\nu^T M_\nu U_\nu =\text{diag}(m_{\nu 1},m_{\nu 2},m_{\nu 3}).
\end{equation}
The entries in this matrix are complex numbers, so the physical masses are their absolute
values.
The mixing matrices are, by definition,
\begin{equation} \label{eq:6}
V_{CKM}=U^\dagger_{uL}U_{dL},\quad V_{PMNS}=U^\dagger_{eL}U_\nu K.
\end{equation}
Where K is defined as the matrix that take out the phases of the diagonal  mass matrix,
\begin{equation}
\text{diag}(m_{\nu 1},m_{\nu 2},m_{\nu 3})=K^\dagger \text{diag}(|m_{\nu 1}|,|m_{\nu 2}|,|m_{\nu 3}|)K^\dagger.
\end{equation}
A further reduction of the number of parameters in the leptonic sector may be achieved by
means of an Abelian $Z_2$ symmetry. A possible set of charge assignments of $Z_2$, compatible with
the experimental data on masses and mixings in the leptonic sector is given in Table I.
\begin{table}[h!]  
\centering
\begin{tabular}{|r|l|}
  \hline
$-$&$+$\\
  \hline
$H_S,\nu_{3R}$&$H_I,L_3,L_I,e_{eR},e_{IR}.\nu_{IR}$\\
  \hline
\end{tabular}
\caption{Z2 assignment in the leptonic sector.} \label{tab:1}
\end{table}
The $Z_2$ assignments forbid the following Yukawa couplings
\begin{equation}
Y_1^e=Y_3^e=Y_1^\nu=Y_5^\nu.
\end{equation} 
Therefore, the corresponding entries in the mass matrices vanish.

\subsection{Mass matrix for the charged leptons}
Under these assumptions, the mass matrix of the charged leptons takes the form 
\begin{equation}
M_e=m_\tau
\begin{pmatrix}
\tilde{\mu_2}&\tilde{\mu_2}&\tilde{\mu_5}\\
\tilde{\mu_2}&-\tilde{\mu_2}&\tilde{\mu_5}\\
\tilde{\mu_4}&\tilde{\mu_4}&0\\
\end{pmatrix}.
\end{equation} 
The unitary matrix $U_{eL}$ that enters in the definition of the mixing matrix, $V_{PMNS}$, is calculated from
 \begin{equation}
U^\dagger_{eL}M_e M_e^\dagger U_{eL}=\text{diag}(m^2_e,m^2_\mu,m^2_\tau),
\end{equation} 
where  $m_3,m_\mu$ and $m_\tau$ are the masses of the charged leptons, and 
\begin{equation}
M_e M_e^\tau=m^2_\tau
\begin{pmatrix}
2|\tilde{\mu_2}|^2&|\tilde{\mu_5}|^2&2|\tilde{\mu_2}||\tilde{\mu_4}|e^{-i \delta_e}\\
|\tilde{\mu_5}|^2&2|\tilde{\mu_2}|^2&0\\
2|\tilde{\mu_2}||\tilde{\mu_4}|e^{i \delta_e}&0&2|\tilde{\mu_4}|^2\\
\end{pmatrix}.
\end{equation} 
Notice that this matrix only has one phase factor. The parameters $|\tilde{\mu_2}|, |\tilde{\mu_4}|$ and $|\tilde{\mu_5}|$ may readily be expressed in terms of the charged lepton masses. From the invariants of $M_e M^\dagger_e$, we get the set of equations \cite{Mondragon:2007af} 
\begin{equation}
Tr(M_e M^\dagger_e)=m^2_e+m^2_\mu+m^2_\tau=m^2_\tau[4|\tilde{\mu_2}|^2+2(|\tilde{\mu_4}|^2+|\tilde{\mu_5}|^2)]
\end{equation}
\begin{align}
\xi(M_e M^\dagger_e)=&m^2_\tau(m^2_e+m^2_\mu)+m_e^2 m_\mu^2\\
=&4m^4_\tau[|\tilde{\mu_2}|^2+|\tilde{\mu_2}|^2(|\tilde{\mu_4}|^2+|\tilde{\mu_5}|^2)+|\tilde{\mu_4}|^2|\tilde{\mu_5}|^2]\\
\end{align}
\begin{equation}
\det(M_eM_e^\dagger)=m_e^2m_\mu^2m_\tau^2=4 m^6_\tau|\tilde{\mu_2}|^2|\tilde{\mu_4}|^2|\tilde{\mu_5}|^2,
\end{equation}
where $\xi(M_eM_e^\dagger)=\frac{1}{2}[Tr(*M_eM_e^\dagger))^2-Tr((M_eM_e^\dagger)^2)]$.
Solving these equations for $|\tilde{\mu_2}|,|\tilde{\mu_4}|$ and $|\tilde{\mu_5}|$, we obtain
\begin{equation}
|\tilde{\mu_2}|^2=\frac{1}{2}\frac{m_e^2+m^2_\mu}{m^2_\tau}-\frac{m_e^2m_\mu^2}{m^2_\tau(m^2_e+m^2_\mu)}+\beta.
\end{equation}
In this expression, $\beta$ is the smallest solution of the equation
\begin{align}
\beta^3-\frac{1}{2}(1-2y+6\frac{x}{y})\beta^2-\frac{1}{4}(y-y^2-4\frac{z}{y}+7z-12\frac{z^2}{y^2})\beta-\\
\frac{1}{8}yz-\frac{1}{2}\frac{z^2}{y^2}+\frac{3}{4}\frac{z^2}{y}-\frac{z^3}{y^3}=0\\
\end{align}
where $y=(m_e^2+m_\mu^2)/ m_\tau$ and $z=\mu^2_\mu\mu^2_e / \mu^4_\tau$.\\
An estimation of $\beta$ at good order of magnitude is obtained from \cite{Carone:1995xw}
\begin{equation}
\beta \simeq -\frac{m_\mu^2 m_e^2}{2m^2_\tau(m^2_\tau-(m^2_\tau+m^2_e))}.
\end{equation}
The parameters $|\tilde{\mu_4}|^2$ and $|\tilde{\mu_5}|^2$ are in terms of $|\tilde{\mu_2}|^2$,
\begin{align}
|\tilde{\mu_{4,5}}|^2=&\frac{1}{4}(1-\frac{m^2_\mu+m^2_e}{m_\tau^2}+4\frac{m_e^2m^2_\mu}{m_\tau^2(m_e^2+m^2_\mu})-\beta)\\
&\pm\frac{1}{4}(\sqrt{(1-\frac{m^2_\mu+m^2_e}{m_\tau^2}+4\frac{m_e^2m^2_\mu}{m_\tau^2(m_e^2+m^2_\mu})-\beta)^2}-\frac{m_\mu^2 m_e^2}{m^4_\tau}\frac{1}{|\tilde{\mu_2}|^2})\\
\end{align}
Once $M_eM_e^\dagger$ has been reparametrized in terms of the charged
lepton masses, it is straightforward to compute $U_{eL}$ also as a
function of the lepton masses. Here  we will write the result to order $(m_\mu m_e /m^2_\tau)^2$
and $x^4=(m_e/m_\tau)^4$
\begin{equation}
M_e\simeq m_\tau
\begin{pmatrix}
\frac{1}{\sqrt{2}}\frac{\tilde{m_\mu}}{\sqrt{1+x^2}}&\frac{1}{\sqrt{2}}\frac{\tilde{m_\mu}}{\sqrt{1+x^2}}&\frac{1}{\sqrt{2}}\frac{\sqrt{1+x^2-\tilde{m_\mu^2}}}{\sqrt{1+x^2}}\\
\frac{1}{\sqrt{2}}\frac{\tilde{m_\mu}}{\sqrt{1+x^2}}&-\frac{1}{\sqrt{2}}\frac{\tilde{m_\mu}}{\sqrt{1+x^2}}&\frac{1}{\sqrt{2}}\frac{\sqrt{1+x^2-\tilde{m_\mu^2}}}{\sqrt{1+x^2}}\\
\frac{\tilde{m_e}(1+x^2)}{\sqrt{1+x^2-\tilde{m_\mu^2}}}e^{i\delta _e}&\frac{\tilde{m_e}(1+x^2)}{\sqrt{1+x^2-\tilde{m_\mu^2}}}e^{i\delta _e}&0\\
\end{pmatrix}.
\end{equation}
The unitary matrix $U_{eL}$ that diagonalizes $M_eM_e^\dagger$ and
enters in the definition of the neutrino mixing matrix $V_{VPMNS}$,
equation (\ref{eq:6}), is
\begin{equation}
U_{eL}\simeq
\begin{pmatrix}
1&0&0\\
0&1&0\\
0&0&e^{i\delta_e}\\
\end{pmatrix}
\begin{pmatrix}
O_{11}&-O_{12}&O_{13}\\
-O_{21}&O_{22}&O_{23}\\
-O_{31}&-O_{32}&O_{33}\\
\end{pmatrix},
\end{equation}
where 
\begin{equation}
U_{eL}\simeq
\begin{pmatrix}
O_{11}&-O_{12}&O_{13}\\
-O_{21}&O_{22}&O_{23}\\
-O_{31}&-O_{32}&O_{33}\\
\end{pmatrix}=
\end{equation}
\begin{equation}
\begin{pmatrix}
\frac{1}{\sqrt{2}}x\frac{(1+2\tilde{m^2_\mu}+x^2+\tilde{m^4_\mu}+2\tilde{m^2}_e)}{\sqrt{1+\tilde{m_\mu^2}+5x^2-\tilde{m_\mu^4}-\tilde{m^6_\mu}+\tilde{m_e^{12}}+12x^4} }&-\frac{1}{\sqrt{2}}\frac{(1-2\tilde{m^2_\mu}+\tilde{m^4_\mu}-2\tilde{m^2}_e)}{\sqrt{1-\tilde{m_\mu^2}+x^2+6\tilde{m_\mu^4}-4\tilde{m^6_\mu}+5\tilde{m_e^{12}}} }&\frac{1}{\sqrt{2}}\\
-\frac{1}{\sqrt{2}}x\frac{(1+4x^2-\tilde{m^4_\mu}-2\tilde{m^2}_e)}{\sqrt{1+\tilde{m_\mu^2}+5x^2-\tilde{m_\mu^4}-\tilde{m^6_\mu}+\tilde{m_e^{12}}+12x^4} }&\frac{1}{\sqrt{2}}\frac{(1-2\tilde{m^2_\mu}+\tilde{m^4_\mu})}{\sqrt{1-\tilde{m_\mu^2}+x^2+6\tilde{m_\mu^4}-4\tilde{m^6_\mu}+5\tilde{m_e^{12}}} }&\frac{1}{\sqrt{2}}\\
-\frac{\sqrt{1+2x^2-\tilde{m^2_\mu}-\tilde{m^2_e}}(1+\tilde{m^2_\mu}+x^2-2\tilde{m^2}_e)}{\sqrt{1+\tilde{m_\mu^2}+5x^2-\tilde{m_\mu^4}-\tilde{m^6_\mu}+\tilde{m_e^{12}}+12x^4} }&-x\frac{\sqrt{1+2x^2-\tilde{m_\mu^2}-\tilde{m_e^2} }(1+x^2-\tilde{m^2_\mu}-2\tilde{m^2}_e)}{\sqrt{1-\tilde{m_\mu^2}+x^2+6\tilde{m_\mu^4}-4\tilde{m^6_\mu}+5\tilde{m_e^{12}}} }&\tilde{m_e}\tilde{m_\mu}\frac{\sqrt{1+x^2}}{\sqrt{1+x^2-\tilde{m^2_\mu}}}\\
\end{pmatrix}=
\end{equation}
and where $\tilde{m_\mu}=m_\mu/m_\tau,\tilde{m_e}=m_e/m_\tau$ and
$x=m_e/m_\mu$.

\subsection{The mass matrix of the neutrinos}

With the $Z_2$ selection rule (Table \ref{tab:1}), the mass matrix of the Dirac neutrinos takes the form
\begin{equation}
M_{\nu D}=
\begin{pmatrix}
\mu^\nu_2&\mu^\nu_2&0\\
\mu^\nu_2&-\mu^\nu_2&0\\
\mu^\nu_4&\mu^\nu_4&\mu^\nu_3
\end{pmatrix}
\end{equation}
Then, the mass matrix for the left-handed Majorana neutrinos is obtained from the see-saw mechanism,
\begin{equation}
M_{\nu}=M_{\nu D} \tilde{M}^{-1} (M_{\nu D})^T=
\begin{pmatrix}
(\frac{1}{M_1}+\frac{1}{M_2})\mu_2^2&(\frac{1}{M_1}-\frac{1}{M_2})\mu_2^2&(\frac{1}{M_1}+\frac{1}{M_2})\mu_2\mu_4\\
(\frac{1}{M_1}-\frac{1}{M_2})\mu_2^2&(\frac{1}{M_1}+\frac{1}{M_2})\mu_2^2&(\frac{1}{M_1}-\frac{1}{M_2})\mu_2\mu_4\\
(\frac{1}{M_1}+\frac{1}{M_2})\mu_2\mu_4&(\frac{1}{M_1}-\frac{1}{M_2})\mu_2\mu_4&\frac{\mu_4^2}{M_2}+\frac{\mu_3^2}{M_3}
\\
\end{pmatrix} ,
\end{equation}
where $M_i$ are the right handed neutrino masses appearing in eq.~(\ref{eq:5}).\\
The non-Hermitian, complex, symmetric neutrino mass matrix $M_\nu$ may be brought to a
diagonal form by a bi-unitary transformation, as
\begin{equation} \label{eq:2}
U^T_\nu M_\nu U_\nu=\text{diag}(m_{\nu 1} e^{i \phi _1},m_{\nu 2} e^{i \phi _2},m_{\nu 3} e^{i \phi _3})
\end{equation}
Where $U_{\nu}$ is the matrix that diagonalizes the matrix $M_{\nu}$.

\subsubsection{Neutrino matrix with degenerate masses.}

In the case where $M_1=M_2$ the mass matrix is reduced to \cite{Felix:2006pn}
\begin{equation} \label{eq:7}
M_{\nu}=M_{\nu D} \tilde{M}^{-1} (M_{\nu D})^T=
\begin{pmatrix}
(\frac{1}{M_1}+\frac{1}{M_1})\mu_2^2&0&(\frac{1}{M_1}+\frac{1}{M_1})\mu_2\mu_4\\
0&(\frac{1}{M_1}+\frac{1}{M_1})\mu_2^2&0\\
(\frac{1}{M_1}+\frac{1}{M_1})\mu_2\mu_4&0&\frac{\mu_4^2}{M1}+\frac{\mu_3^2}{M3}\\
\end{pmatrix}.
\end{equation}
With this texture is easy to calculate the $U_\nu$ matrix that
diagonalizes $M_\nu^\dagger M_\nu$, 
\begin{equation}
M^\dagger_\nu M_\nu=
\begin{pmatrix}
|A|^2+|B|^2&0&A^*B+B^*D\\
0&|A^2|&0\\
AB^*+BD^*&0&|B|^2+|D|^2
\end{pmatrix}
\end{equation}
with $A=\mu_2 ^2 / M_1 , B=2 \mu_2 \mu_4/M_1 $and $D=2 \mu_1 \mu_2 / M_1+\mu_3 ^2 /M_2$, this matrix is diagonalized by 
\begin{equation} \label{eq:1}
U_\nu=
\begin{pmatrix}
1&0&0\\
0&1&0\\
0&0&e^{i\delta_\nu}\\
\end{pmatrix}
\begin{pmatrix}
\cos \eta & \sin \eta & 0 \\
0&0&1\\
-\sin\eta&\cos\eta&0\\
\end{pmatrix}.
\end{equation}
If we require that the defining equation (\ref{eq:2}) be satisfied as an identity, we get the following set
of equations:
\begin{align}
2(\mu_2^\nu)^2/M1=&m_{\nu3 },\\
2(\mu_2^\nu)^2/M1=&m_{\nu1 }\cos^2\eta+m_{\nu2 }\sin^2\eta,\\
2(\mu_2^\nu)(\mu_4^\nu)/M1=&\sin\eta \cos\eta(m_{\nu2}-m_{\nu1})e^{i\delta_nu},\\
2(\mu_2^\nu)^2/M1=&m_{\nu3 },\\ \label{eq:8}
2(\mu_2^\nu)(\mu_4^\nu)/M1=&\sin\eta \cos\eta(m_{\nu2}-m_{\nu1})e^{i\delta_nu},\\ \label{eq:9}
2(\mu_4^\nu)^2/M1+2(\mu_3^\nu)^2/M3=&(m_{\nu1}\sin^2\eta+m_{\nu2}\cos^2\eta)e^{-2i\delta_nu}.\\ \label{eq:10}
\end{align}
Solving these equations for $\sin \eta$ and $\cos \eta$, we find
\begin{equation}
\sin^2\eta=\frac{m_{\nu 3}-m_{\nu 1}}{m_{\nu 2}-m_{\nu 1}} \quad \cos^2\eta=\frac{m_{\nu 2}-m_{\nu 3}}{m_{\nu 2}-m_{\nu 1}}.
\end{equation}
The unitarity of $U_\nu$ constrains $\sin \eta$ to be real and thus $|\sin \eta| \leq 1$, this condition fixes the phases $\phi_1$ and $\phi_2$ as
\begin{equation}
|m_{\nu 1}|\sin{\phi_1}=|m_{\nu 2}|\sin{\phi_2}=|m_{\nu 3}|\sin{\phi_3}.
\end{equation}
The real phase $\delta_\nu$ appearing in eq. (\ref{eq:1}) is not constrained by the unitarity of $U_\nu$.
Therefore the $U_\nu$ matrix is,
\begin{equation}
U_\nu=
\begin{pmatrix}
1&0&0\\
0&1&0\\
0&0&e^{i\delta_nu}\\
\end{pmatrix}
\begin{pmatrix}
\sqrt{\frac{m_{\nu2}-m_{\nu3}}{m_{\nu2}-m_{\nu1} }}&\sqrt{\frac{m_{\nu3}-m_{\nu1}}{m_{\nu2}-m_{\nu1} }}& 0 \\
0&0&1\\
-\sqrt{\frac{m_{\nu3}-m_{\nu1}}{m_{\nu2}-m_{\nu1} }}&\sqrt{\frac{m_{\nu2}-m_{\nu3}}{m_{\nu2}-m_{\nu1} }}& 0 \\
\end{pmatrix}.
\end{equation}
Now, the mass matrix of the Majorana neutrinos, $M_\nu$, may be written in terms of the neutrino
masses; from (\ref{eq:7}) and (\ref{eq:8},\ref{eq:9},\ref{eq:10}), we get
\begin{equation}
M_{\nu}=
\begin{pmatrix}
m_{\nu3}&0&\sqrt{(m_{\nu3}-m_{\nu1})(m_{\nu2}-m_{\nu3})}e^{-i\delta_\nu}\\
0&m_{\nu3}&0\\
\sqrt{(m_{\nu3}-m_{\nu1})(m_{\nu2}-m_{\nu3})}e^{-i\delta_\nu}&0&(m_{\nu1}+m_{\nu2}-m_{\nu3})e^{-2\delta_\nu}\\
\end{pmatrix}
\end{equation}
The only free parameters in these matrices, other than the neutrino
masses, are the phase $\phi_\nu$, implicit in $m_{\nu1},m_{\nu2}$ and $m_{\nu3}$, and the Dirac phase $\delta_\nu$.

Therefore, the theoretical mixing matrix $V_{PMNS}$ , is
given by
\begin{equation}
 V^{th}_{PMNS}=\begin{pmatrix}
 O_{11} \cos\eta + O_{31} \sin\eta e^{i\delta} & O_{11} \sin\eta - O_{31} \cos\eta e^{i\delta} &-O_{21} \\
 -O_{12} \cos\eta + O_{32} \sin\eta e^{i\delta} & -O_{12} \sin\eta - O_{32} \cos\eta e^{i\delta} &O_{22}\\
O_{13} \cos\eta - O_{33} \sin\eta e^{i\delta} & O_{13} \sin\eta + O_{33} \cos\eta e^{i\delta} &O_{23}\\
\end{pmatrix}
 \times K.
\end{equation}
To obtain the expressions for the mixing angles we need to match the
theoretical and PDG expressions for the $V_{PMNS}$ matrix 
\begin{equation}
|V_{PMNS}^{th}|=|V_{PMNS}^{PDG}|
\end{equation}
meaning $|V_{ij}^{th}|=|V_{ij}^{PDG}|$.
The standard parametrization of the Particle Data Group is 
\begin{equation}
V_{PMNS}=
\begin{pmatrix}
c_{12}c_{13}&s_{12}c_{13}&s_{13}e^{-i \delta_{CP}} \\
-s_{12}c_{23}-c_{12}s_{23}s_{13}e^{i \delta_{CP}}&c_{12}c_{23}-s_{12}s_{23}s_{13}e^{i \delta_{CP}}&s_{23}c_{13}\\
s_{12}s_{23}-c_{12}c_{23}s_{13}e^{i \delta_{CP}}&-c_{12}s_{23}-s_{12}c_{23}s_{13}e^{i \delta_{CP}}&c_{23}c_{13}\\
\end{pmatrix}.
\end{equation}
We can straightforwardly read the equation for the mixing angles with
\begin{equation}
|\sin_{\theta_{13}}|=|O_{21}|\simeq\frac{1}{\sqrt{2}}x\frac{1+4x^2-\tilde{m_\mu^4}}{\sqrt{1+\tilde{m_\mu^2}+5x^2-\tilde{m_\mu^4}}}~,
\end{equation}
\begin{equation}
|\sin_{\theta_{23}}|=\frac{|O_{22}|}{\sqrt{1-O^2_{21} } }\simeq \frac{1}{\sqrt{2}} \frac{1-2\tilde{m_\mu^2}+\tilde{m_\mu^4}}{\sqrt{1-4\tilde{m_\mu^2}+x^2+6\tilde{m_\mu^4}}}~,
\end{equation}
and
\begin{equation}
\tan_{\theta_{12}}=\frac{O_{11}\sin_\eta-O_{31}\cos_\eta}{O_{31}\sin_\eta+O_{11}\cos_\eta} 
\end{equation}
\begin{equation}
\simeq -\sqrt{\frac{m_{\nu2}-m_{\nu3} } {m_{\nu3}-m_{\nu1}} }\times (\frac{\sqrt{1+2x^2-\tilde{m^2_{\mu} } }(1+\tilde{m_\mu^2}+x^2)-\frac{1}{\sqrt{2}}x(1+2\tilde{m_\mu^2}+4x^2)\sqrt{\frac{m_{\nu3}-m_{\nu1} } {m_{\nu2}-m_{\nu3}} } }{\sqrt{1+2x^2-\tilde{m^2_{\mu} } }(1+\tilde{m_\mu^2}+x^2)+\frac{1}{\sqrt{2}}x(1+2\tilde{m_\mu^2}+4x^2)\sqrt{\frac{m_{\nu2}-m_{\nu3} } {m_{\nu3}-m_{\nu1}} } }).
\end{equation}
We can express $\tan \theta_{12}$ in terms of the differences of the
square of the masses as 
\begin{equation} \label{eq:11}
\tan^2\theta_{12}=\frac{(\Delta m^2_{12}+\Delta m^2_{13}+|m_{\nu 3}|^2 \cos^2\phi_{nu})^{1/2}-|m_{\nu 3}||\cos\phi_{nu}|}    {(\Delta m^2_{13}+|m_{\nu 3}|^2 \cos^2\phi_{nu})^{1/2}+|m_{\nu 3}||\cos\phi_{nu}|}
\end{equation}
where $\Delta m^2_{ij}=m_{\nu i}^2-m_{\nu j}^2$.\\
We can use the experimental values of the masses of the charged leptons and the differences of the square of the masses to fit the mixing angles, 
\begin{equation}
(\sin^2\theta_{13})^{th}=1.1\times10^{-5},\quad (\sin^2\theta_{13})^{xp}=2.19^{+0.12}_{-0.12}\times 10^{-2},
\end{equation}
and
\begin{equation}
(\sin^2\theta_{23})^{th}=.499,\quad (\sin^2\theta_{23})^{xp}=.5^{+0.05}_{-0.05}.
\end{equation}
From expression (\ref{eq:11}), we may readily derive expressions for
the neutrino masses in terms of $\tan\theta_{12}, \phi_\nu$ and the differences of the squared masses,
%
%
\begin{equation}
|m_3|=\frac{\sqrt{\Delta m_{13}^2}}{2\tan\theta_{12}\cos\phi_\nu}\frac{1-\tan^4\theta_{12}+r^2}{(1+\tan^2\theta_{12})(1+\tan^2\theta_{12}+r^2)}~,
\end{equation}
\begin{equation}
|m_1|=\sqrt{|m_{\nu 3}|^2+\Delta m_{13}^2}~,
\end{equation}
\begin{equation}
|m_2|=\sqrt{|m_{\nu 3}|^2+\Delta m_{13}^2(1+r^2)}~,
\end{equation}
here $r^2 = \Delta m^2_{12} / \Delta m^2_{13} \approx 3 \times 10^{-2}
$. This implies an inverted neutrino mass spectrum $|m_{\nu 3} |
< |m_{\nu 1} | < |m_{\nu 2} |$. 
As $r^ 2 << 1$, the sum of the neutrino masses is
\begin{equation}
\sum^{3}_{i=1} | m_{\nu_i} | \approx \frac{\Delta m^2_{13}}{2\cos \phi_\nu \tan \theta_{12}} (1 + 2 \sqrt{1 + 2 \tan^2 \theta_{12} (2 \cos^2 \phi_\nu - 1) + \tan^4 \theta_{12}} - \tan^2 \theta_{12}) .
\end{equation}
The most restrictive cosmological upper bound \cite{Ade:2013zuv} for this sum is 
\begin{equation} \label{eq:12}
\sum|m_\nu | \leq 0.23eV~.                                   
\end{equation}
This upper bound and the experimentally determined values of $\tan \theta_{12}$ and $\Delta m^2_{i,j}$, give a lower bound for 
\begin{equation}
\cos \phi_\nu \geq 0.55                                         
\end{equation}
or $0 \leq \phi_\nu \leq 57^\circ $.
We can use again equation (\ref{eq:11}) to set the best value of $\phi$, we find that with $\phi=50^\circ$ we get, 
\begin{equation}
\tan \theta_{12}=0.665288
\end{equation}
Hence, setting $\phi_\nu = 50^\circ$ in
our formula, we find
\begin{equation}
m_{\nu 1} = 0.052 ~eV,\quad   m_{\nu 2} = 0.053 ~eV,\quad m_{\nu 3} = 0.019 ~eV.
\end{equation}                                                
The computed sum of the neutrino masses is
\begin{equation} 
(\sum^3_{i=1} |m_{\nu i} |)^{th}      = 0.168508 ~eV,                           
\end{equation} 
below the cosmological upper bound given in eq.~(\ref{eq:12}), as
expected.
The above value of $\phi$ is in  agreement with the requirements for
leptogenesis, as we will show in section 3.
One of the successes of the S3-3H model  has been to predict  an angle
$\theta_{13}$ different from zero, as well a very accurate angles
$\theta_{12}$ and $\theta_{23}$.  Nevertheless new experimental
results have shown that the angle $\theta_{13}$ is greater than the
model predicts with degenerate right-handed neutrino masses. This is
the major reason to extend the model further, to the non-degenerate
case \cite{Canales:2012dr},  and  where the angles fit the experimental value.  

\subsubsection{The mass matrix of the neutrinos without degeneration}
In a more extensive analysis than \cite{Canales:2012dr}, we continue to study the case where the RHN masses are non-degenerate.
The effective neutrino mass matrix
$m_\nu$ is,
\begin{equation}
M_{\nu}=M_{\nu D} \tilde{M}^{-1} (M_{\nu D})^T=
\begin{pmatrix}
(\frac{1}{M_1}+\frac{1}{M_2})\mu_2^2&(\frac{1}{M_1}-\frac{1}{M_2})\mu_2^2&(\frac{1}{M_1}+\frac{1}{M_2})\mu_2\mu_4\\
(\frac{1}{M_1}-\frac{1}{M_2})\mu_2^2&(\frac{1}{M_1}+\frac{1}{M_2})\mu_2^2&(\frac{1}{M_1}-\frac{1}{M_2})\mu_2\mu_4\\
(\frac{1}{M_1}+\frac{1}{M_2})\mu_2\mu_4&(\frac{1}{M_1}-\frac{1}{M_2})\mu_2\mu_4&\frac{\mu_4^2}{M2}+\frac{\mu_3^2}{M3}\\
\end{pmatrix}.
\end{equation}
We are going to assume that the phases of the $\mu_3$ and $\mu_4$
terms are aligned, therefore we can write $M_\nu$ in polar form
$M_\nu=P \widetilde{M_\nu} P$, with $\widetilde{M_\nu}$, real and
$P=\text{diag}(e^{-i \theta_\mu 2},e^{-i \theta_{\mu 2}},e^{i(\frac{ca}{2}
  -\theta{_{\mu 4}})})$ In this way $\widetilde{M_\nu}$ can be
expressed in terms of a matrix with two texture zeros class I as:
\begin{equation}
\widetilde{M_\nu}=\mu_0 \mathbb{I}_{3\times 3}+M_\nu'
\end{equation}
with $\mu_0=\frac{2 |\mu_2|^2}{M1}$.\\
Therefore, the $U_\nu$ matrix is $P^\dagger U_1$ where $U_1$ is the matrix that diagonalizes $M'_\nu$.
We can take a rotation $u_{\pi /4}$,
\begin{equation}
u_{\pi /4}=\frac{1}{\sqrt{2}}
\begin{pmatrix}
1&1&0\\
-1&1&0\\
0&0&\sqrt{2}\
\end{pmatrix},
\end{equation}
to the $M'_\nu$ matrix, 
\begin{equation}
u_{\pi /4}^T M'_\nu u_{\pi /4}=na
\begin{pmatrix}
0&0&1\\
0&\sqrt{2}\psi (\psi_2-1)&\psi_2\\
1&\psi_2&\mu_c
\end{pmatrix}
\end{equation}
with $na=M_2/(\sqrt{2}|\mu_2||\mu_4|)$, $\psi=\frac{\sqrt{2}|\mu_2|}{|\mu_4|}$, $\psi_2=\frac{M_2}{M_1}$ and $\mu_c=\frac{\mu_4^2}{M1}+\frac{\mu_3^2}{M3}-\mu_0$.
The Matrix that diagonalizes $u_{\pi /4}^T M'_\nu u_{\pi /4}$ is 
\begin{equation}
U_2=
\begin{pmatrix}
\psi_2 n_1 &\psi_2 n_2 &\psi_2 n_3 \\
-(1+\mu_c \lambda_1-\lambda^2_1) n_1&-(1+\mu_c \lambda_2-\lambda^2_2) n_2&-(1+\mu_c \lambda_3-\lambda^2_3) n_3\\
\psi_2 \lambda_1 n_1&\psi_2 \lambda_2 n_2&\psi_2 \lambda_3 n_3\
\end{pmatrix}.
\end{equation}
where $n_i$ is a normalization factor and $\lambda_i$ is the i-th eigenvalue of  $M'_\nu$.With $U_1=u_{\pi/4} U_2$, therefore

\begin{equation}
U_\nu=
\begin{pmatrix}
O'_{11}&O'_{12}&O'_{13}\\
O'_{21}&O'_{22}&O'_{23}\\
O'_{31}&O'_{32}&O'_{33}\\
\end{pmatrix},
\end{equation}
\begin{equation}
\begin{pmatrix}
n_1 (\psi_2+ f_1)&n_2 (\psi_2+ f_2)&n_3 (\psi_2+ f_3)\\
n_1 (-\psi_2+ f_1)&n_2 (-\psi_2+ f_2)&n_3 (-\psi_2+ f_3)\\
\psi_2 n_1 \lambda_1&\psi_2 n_2 \lambda_2&\psi_2 n_3 \lambda_3\\
\end{pmatrix},
\end{equation}
where $f_i=(-1-\mu_c \lambda_i+\lambda^2_i)$. 
In the same way as in  the degenerate scenario we have
\begin{equation}
|V_{PMNS}^{th}|=|V_{PMNS}^{PDG}|~,
\end{equation}
in terms of the mixing angles with
\begin{equation}
s_{13}=O'_{13}~, \quad s_{23}=\frac{O'_{23}}{\sqrt{1-O_{13}^{'2}}}~, \quad s_{12}=\frac{O'_{12}}{\sqrt{1-O_{13}^{'2}}}~.
\end{equation}
In the non-degenerate scenario we have three free parameters ($\psi,
\quad \psi_2, \quad \mu_c$) for the neutrino matrix. In this model the
PMNS matrix may be obtained numerically. We have used the following values for the masses given in \cite{Olive:2016xmw}
\begin{align}
m_e =& 0.5109989461 \pm 0.0000000031 ~MeV,\\
m_\mu =& 105.6583745 \pm 0.0000024 ~MeV,\\
m_\tau =& 1776.86 \pm 0.12 ~MeV.\\
\end{align}
In order to obtain the numerical values for the three free parameters 
we perform a $\chi^2$ analysis on the parameter space to find their
best fit points
\begin{equation}
\chi^2=\frac{(\sin(\theta_{12})^2-\sin(\theta_{12})^2)^2}{\sigma^2_{\sin\theta_{12}^2}}+\frac{(\sin(\theta_{12})^2-\sin(\theta_{12})^2)^2}{\sigma^2_{\sin\theta_{12}^2}}+\frac{(\sin(\theta_{12})^2-\sin(\theta_{12})^2)^2}{\sigma^2_{\sin\theta_{12}^2}}
\end{equation}
where we have taken the following experimental values for the $V_{PMNS}$ elements \cite{Olive:2016xmw}
\begin{equation}
\sin^2_{\theta_{12}}= 0.304 \pm 0.014 ,\quad
\sin^2_{\theta_{23}} = 0.50 \pm 0.05 ,\quad
\sin^2_{\theta_{13}}= (2.19 \pm 0.12) \times 10^{-2}~.
\end{equation}
The best values for the free parameters are thus found to be
\begin{equation}
\psi_2 = 1.1431~,~~
\psi = 1.3091~,~~
\mu_c = 1.6502~eV,
\end{equation}
at one sigma C.L. with $\chi^2=3.74\times10^{-15}$ as the minimal value. These correspond to the following mixing angles
\begin{equation}
\begin{split}
\sin(\theta_{12})^2=&0.3039, \quad
\sin(\theta_{23})^2=0.4999, \\
& \sin(\theta_{13})^2=0.0218.
\end{split}
\end{equation}
\begin{figure}[H]
  Fig \ref{fig:parspace} shows the values for ($\psi, \psi_2, \mu_c$) resulting from the $\chi^2$ analysis.  This is part of
  a more general analysis on the neutrino sector of this
  model \cite{inprep}.

\begin{center}
\textbf{Parameter space}\par\medskip
\end{center}

\begin{subfigure}{0.5\textwidth}
\includegraphics[width=1\linewidth]{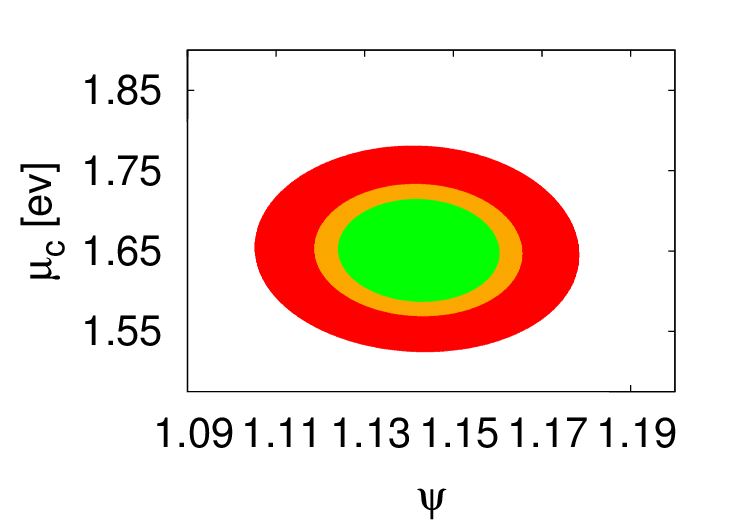}
\end{subfigure}
\begin{subfigure}{0.5\textwidth}
\includegraphics[width=1\linewidth]{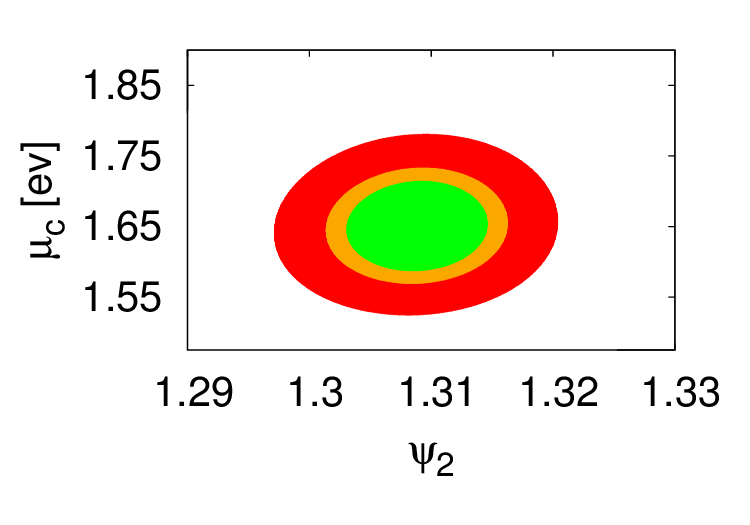}
\end{subfigure}
\end{figure}
\begin{figure}[H]
\begin{center}
\includegraphics[width=.5\linewidth]{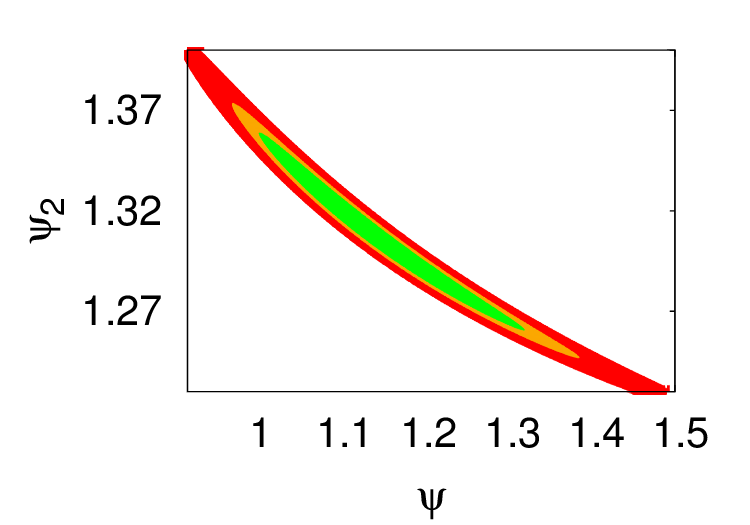}
\end{center}
\caption{Plot of the free parameter space for the non-degenerate case,  where `green' is one,
  `orange' two, and `red' is three $\sigma$ agreement with the experimental value.}\label{fig:parspace}
\end{figure}

\section{Leptogenesis in an S3-3H model}
\begin{figure}[h]
\includegraphics[width=4.5in,height=1.4in,angle=0]{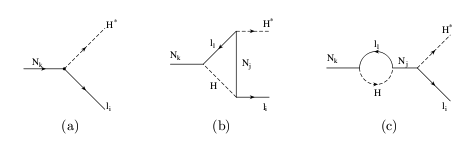}
\caption{Feynman diagrams up to second order. a) Tree level b) Second Order c) Self Energy diagram }
\label{fig1}
\end{figure}

The Yukawa couplings of the neutrinos allow the decay of the right-handed neutrinos into the left-handed ones.
\begin{equation}
-Y^{{\nu}}_1\overline{L_I}(i\sigma_2H^{\ast}_S{\nu}_{IR}).
\end{equation}
As shown in \cite{Chen:2007fv,Davidson:2008bu}, the asymmetry is defined to be
\begin{equation}
\epsilon _1=\frac{\sum_\alpha \Gamma (N_1\rightarrow \ell _\alpha H)-\Gamma (N_1\rightarrow \overline{\ell} _\alpha \overline{H})} {\sum_\alpha \Gamma (N_1\rightarrow \ell _\alpha H) +\Gamma (N_1\rightarrow \overline{\ell} _\alpha \overline{H})} .
\end{equation}
where $\Gamma$ is the decay rate, and $N_1$ is the decaying right-handed neutrino.\\
The possible decays up to tree level are shown in \ref{fig1},
\\
The asymmetry generated by these decays are,
\begin{equation}
 \epsilon\simeq-\frac{3}{8\pi}\frac{1}{(h_{\nu}h^{\dagger}_{\nu})}\sum_{i=2,3}\text{Im}\{ (h_{\nu}h^{\dagger}_{\nu})^2_{1i}  \}[f(\frac{M^2_i}{M^2_1})+g(\frac{M^2_i}{M^2_1})].
\end{equation}
and the self interactions are
\begin{equation}
f(x)=\sqrt{x}[1-(1+x)\ln(\frac{1+x}{x})]~,
\end{equation}
\begin{equation}
g(x)=\frac{\sqrt{x}}{1-x}~.
\end{equation}
This function depends strongly on the hierarchy of the light neutrino masses. It can lead to
a strong enhancement of the CP asymmetries if the masses $M_2$ and $M_3$ are
nearly degenerate.\\
The relation between the lepton and baryon asymmetry is given through the sphaleron process \cite{Kuzmin:1985mm}
\begin{equation}
Y_B=aY_{B-L}=\frac{a}{a-1}Y_L ~,
\end{equation}
where a is  $a=(8N_f+4N_H)/(22N_f+13N_H)$, $N_f$ is the number of families and $N_H$ is the number of Higgs doublets.\\
We can express the lepton asymmetry in terms of the CP asymmetry 
\begin{equation}
Y_L=\frac{n_L-n_{\bar{L}}}{s}=\kappa\frac{\epsilon_i}{g*}~,
\end{equation}
where g is 110, the number of relativistic degrees of freedom,
$\kappa$ is obtained from solving the Boltzmann equations, and it can be
reparametrized in terms of $K$ defined as the ratio of $\Gamma_1$, the
tree-level decay width of $N_1$ to H the Hubble parameter at
temperature $M_1$, where $k=\Gamma_1/H<1$ describes a process out of
thermal equilibrium and $\kappa<1$ describes the washout effect \cite{Pilaftsis:1998pd,Flanz:1998kr}:
\begin{equation}
\kappa\approx\frac{0.3}{K(\ln(K))^{0.6}} \quad \text{for} \quad 10<K<10^6~,
\end{equation}
\begin{equation}
\kappa\approx\frac{1}{2\sqrt{K^2+9}} \quad \text{for} \quad 0<K<10~.
\end{equation}
The decay width of $N_1$ by the Yukawa interaction at tree level and
Hubble parameter in terms of the temperature T and the Planck scale
$M_{pl}$ are $\Gamma_1=(m^\dagger_D m_D)_{11}M_1/(8\pi v^2)$ and
$H=1.66g^{*1/2}T^2/M_{pl}$ respectively. At temperature $T=M_1$ the
ratio $K$ is
\begin{equation}
K=\frac{M_{pl}}{1.66\sqrt{g*}(8\pi v^2)}\frac{(m^\dagger_D m_D)_{11}}{M_1}~.
\end{equation}

\subsection{Baryon asymmetry in the degenerate scheme}

Putting all the above ingredients together, the asymmetry for the S3-3H model is
\begin{equation}\label{eq:asym}
\epsilon=\frac{\text{Im}[e^{2i\delta^*}M_2 m_3 \frac{\sqrt{M_2(m_2-m_3)(m_3-m_1)}}{\sqrt{m_3}}](f[\frac{M_3^2}{M_1^2}]+g[\frac{M_3^2}{M_1^2}])}{8\pi |M_2 m_3|}~.
\end{equation}
The value of the baryon asymmetry has a dependence on $\phi$ and the
masses of the neutrinos $|m_1|,M_1,|m_2|,M_2,|m_3|,M_3$, where the
masses of the right-handed neutrinos are considered real.  We can
calculate the dependence of the baryon asymmetry on the phase
$\delta$. As can be seen from eq.~(\ref{eq:asym}) the asymmetry is a
periodic function of $\delta$, where the  masses give the
scale of the baryon asymmetry.

\begin{figure}[H]
\begin{center}
\includegraphics[width=0.8\textwidth]{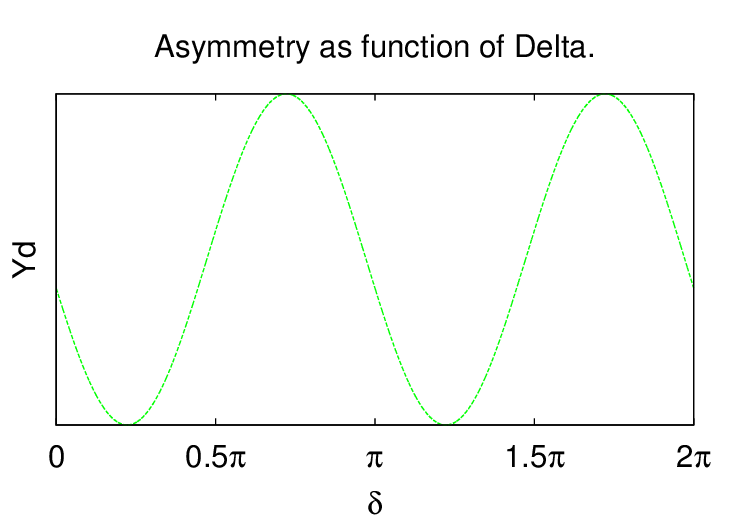}
\caption{Baryon asymmetry dependance on $\delta$.}\label{fig:asymmetry}
\end{center}
\end{figure}
\begin{figure}[H]
\begin{center}
\includegraphics[width=0.8\textwidth]{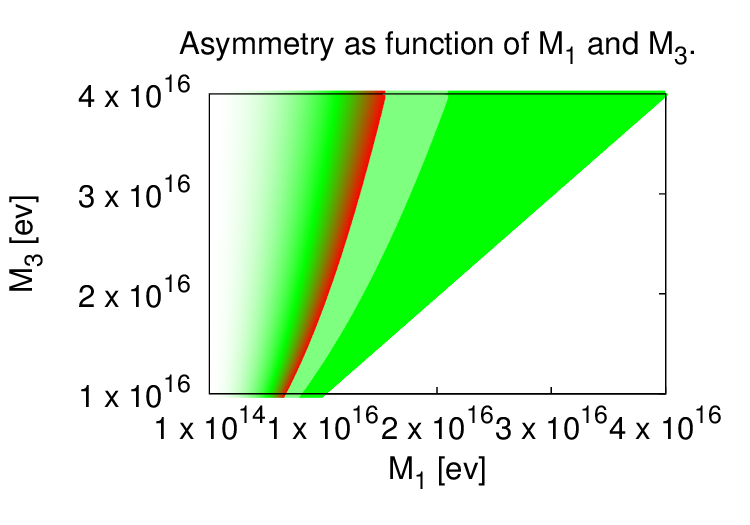}
\caption{Baryon asymmetry dependence on $M_1,M_3$, more asymmetry
  corresponds to darker shades of green. The red region corresponds to
  an excess of asymmetry, with the maximum value of $\delta$.}\label{fig:M1M3}
\end{center}
\end{figure}

The maximum value for the baryon asymmetry is on $\delta=3/4\pi$.  As
figure \ref{fig:asymmetry} shows, the leptogenesis is crucially
dependent on the phase.  The value of the baryon asymmetry is
determined by the masses of the light neutrinos and the ratios of
the right-handed neutrino masses. The see-saw mechanism relates the masses of
the right-handed neutrinos to the light ones, 
making the right-handed neutrino masses bigger than $10^{12}~GeV$, in order to be in agreement with the experimental data.\\
We calculate the asymmetry generated in the best case scenario where
$\delta=3/4\pi$. In this case we can see from fig.~\ref{fig:M1M3} that the masses of the
right-handed neutrinos could be of order of $10^7~GeV$ to produce leptogenesis. The graph also
shows the region of resonant leptogenesis $M_1-M_3\simeq
\frac{1}{2}\Gamma_{N_{1,3}}$, where the asymmetry increases above the
one observed in the Universe, lowering even more the possible mass of
the right-handed neutrinos or the $\delta$ phase.

\subsection{Non-degenerate scenario}

The value of the baryon asymmetry in the non-degenerate case has
dependence on the angles of the $M_d$ matrix $\mu_{2a},~\mu_{3a}$ and
the real masses of the neutrinos $M_1,M_2,M_3$, where $\mu_{ia}$ is
the angle of $\mu_i=|\mu_i|e^{i\mu_{ia}}$.  We can calculate the
dependence of the baryon asymmetry on the phases $\mu_{2a}
,~\mu_{3a}$.  As in the degenerate case the magnitude or scale of the baryon
assymetry is given by the masses.

\begin{figure}[H]
\begin{center}
\includegraphics[width=0.8\textwidth]{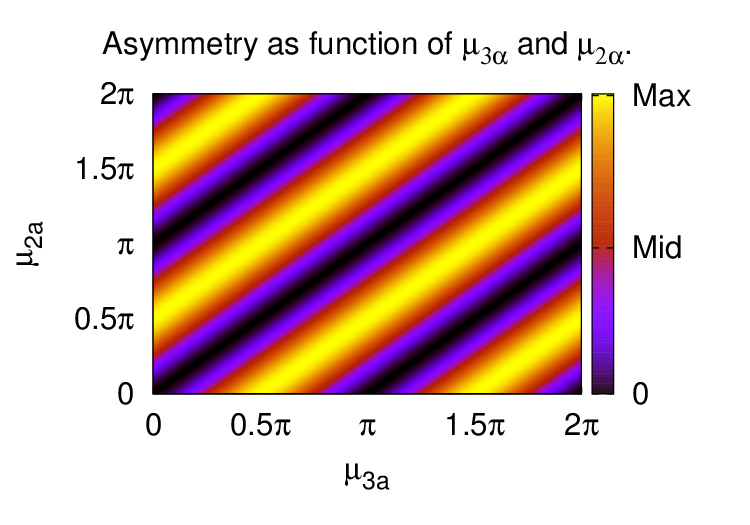}
\caption{Baryon asymmetry dependance on the phases $\mu_{2a},~ \mu_{3a}$}\label{fig:mues}
\end{center}
\end{figure}

\begin{figure}[H]
\begin{center}
\includegraphics[width=0.8\textwidth]{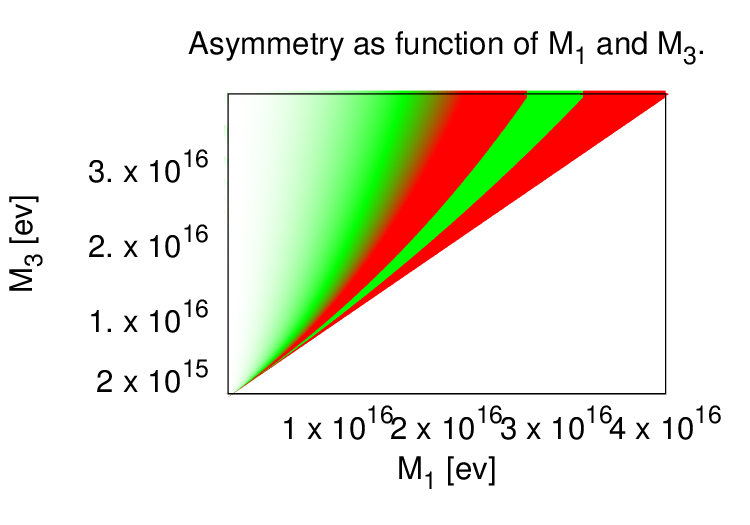}
\caption{Baryon asymmetry depending on the values of $M_1,M_3$. The colour code is similar as in fig.~\ref{fig:M1M3}.}
\label{fig:M1-M3-non}
\end{center}
\end{figure}

The maximum of the asymmetry is achieved in all the lines where
$\mu_{2a}-\mu_{3a}=\pi /2 +n\pi$, where $n$ is any integer, as can be
seen from fig.~\ref{fig:mues}.  Again, this is independent of the
masses of the neutrinos, the masses only fix the scale of the
asymmetry.  Taking the best values of the angles we can see that the
scale of the masses of the right-handed neutrinos can be lower and
that the region of resonant leptogenesis is wider. Therefore, this
gives a wider region in parameter space fulfilling the baryon
asymmetry explanation.  In fig.~\ref{fig:M1-M3-non} we show the baryon
asymmetry dependence on the $M_1$ and $M_3$ masses. The darker shades
of green correspond to more asymmetry, whereas the red regions
correspond to an excess of baryon asymmetry as compared to the one
observed in the Universe, for the maximum value of the phases.

\section{Conclusions}

The minimal S3-3H extension of the SM acommodates well the masses and
mixings of quarks and leptons, and gives naturally a non-zero value
for the neutrino reactor mixing angle.  We re-derived previous results
on the neutrino sector with recent experimental data, taking into
account a new value of the phase $\phi$ to include the angle
$\theta_{12}$ in the model.  In the non-degenerate right-handed
neutrino mass case, we find a new parametrization of the $V_{PMNS}$
matrix and use the experimental values in a $\chi^2$ analysis to fit
the new parameters. We find thus a new region in parameter space where
the model predicts the mixing angles correctly.\\
We then calculated the leptogenesis and the associated baryogenesis in
this model in the case of two right-handed degenerate neutrino masses,
and in the more general case of non-degenerate masses.  We show that
there are regions in parameter space which allow leptogenesis as a
mechanism to solve the observed baryonic asymmetry with right-handed
neutrino masses starting from $10^6~GeV$.

\section*{Acknowledgements} We acknowledge useful discussions with J. Kersten and
  A. Mondrag\'on.  This work is partially supported by a UNAM grant
  PAPIIT IN111115.

\nocite{Frere:2006iz}
\nocite{Davidson:2008bu}
\nocite{Ishimori:2010au}
\nocite{Antusch:2003kp}
\bibliographystyle{h-physrev3}
\bibliography{bibtex,masbiblio}
\end{document}